# Lattice-Tuned Magnetism of $Ru^{4+}$ ($4d^4$) Ions in Single-Crystals of the Layered Honeycomb Ruthenates: $Li_2RuO_3$ and $Na_2RuO_3$

J. C. Wang[1,2,3], J. Terzic[1], T. F. Qi[1], Feng Ye[2,1], S. J. Yuan[1,4], S. Aswartham[1], S. V. Streltsov[5,6], D. I. Khomskii[7], R. K. Kaul[1] and G. Cao[1*]

[1]Center for Advanced Materials, Department of Physics and Astronomy, University of Kentucky, Lexington, KY 40506, USA

[2]Quantum Condensed Matter Division, Oak Ridge National Laboratory, Oak Ridge, Tennessee 37831, USA

[3]Department of Physics, Renmin University of China, Beijing, China

[4]Department of Physics, Shanghai University, Shanghai, China

[5]Institute of Metal Physics, 620041 Ekaterinburg, Russia

[6]Ural Federal University, 620002 Ekaterinburg, Russia

[7]II. Physikalisches Institut, Universitaet zu Koeln, Germany

We synthesize and study single crystals of the layered honeycomb lattice Mott insulators $Na_2RuO_3$ and $Li_2RuO_3$ with magnetic $Ru^{4+}$($4d^4$) ions. The newly found $Na_2RuO_3$ features a nearly ideal honeycomb lattice and orders antiferromagnetically at 30 K. Single-crystals of $Li_2RuO_3$ adopt a honeycomb lattice with either C2/m or more distorted $P2_1/m$ below 300 K, depending on detailed synthesis conditions. We find that $Li_2RuO_3$ in both structures hosts a well-defined magnetic state, in contrast to the singlet ground state found in polycrystalline $Li_2RuO_3$. A phase diagram generated based on our results uncovers a new, direct correlation between the magnetic ground state and basal-plane distortions in the honeycomb ruthenates.

*Introduction*It has been of great interest to study interacting electrons on the honeycomb lattice in various contexts both experimentally (e.g. graphene) and theoretically (e.g the Kitaev model). Studies of honeycomb materials have intensified in recent years **[1-19]**in part because strong spin-orbit coupling (SOC) along with other competing interactions and geometric frustration in the honeycomb iridates $Na_2IrO_3$ and $Li_2IrO_3$favorsa highly anisotropic Kitaev interaction**[20]** that stabilize exotic ground states such as topological spin-liquids**[1]**.It is now experimentally established that $Na_2IrO_3$ exhibits a peculiar zigzag magnetic order at $T_N$=18 K**[5, 14, 15]**, and $Li_2IrO_3$ also orders at $T_N$=15 K but with a different ground stateyet to be defined **[3, 17, 21, 22, 23]**. Indeed, for $(Na_{1-x}Li_x)_2IrO_3$ with $0 \leq x \leq 0.90$, the measured phase diagram demonstrates a dramatic suppression of $T_N$ at intermediate x suggesting that the magnetic order in $Na_2IrO_3$ and $Li_2IrO_3$ is different;however, no spin liquid has been observed thus far**[17]**. Our pursuit of an understanding of the honeycomb iridates has led us to their ruthenate counterparts, $Na_2RuO_3$ and $Li_2RuO_3$. These materials feature $Ru^{4+}(4d^4)$ ions and a weaker or "intermediate strength" SOC (~ 0.16 eV, compared to ~ 0.4 eV for Ir ions) **[24]**. The different d-shell filling and contrasting hierarchy of energy scales between the ruthenates and iridates provide a unique opportunity for a deeper understanding of the fundamental problem of interacting electrons on the honeycomb lattices. The magnetism of $Ru^{4+}$ ions as well as other heavy "$d^4$ ions" (such as $Rh^{5+}(4d^4)$, $Re^{3+}(5d^4)$, $Os^{4+}(5d^4)$ and $Ir^{5+}(5d^4)$)is interesting in their own right, as emphasized recently**[25]**.Materials with heavy$d^4$ ions tend to adopt a low-spin state because larger crystal fields often overpower the Hund's rule coupling. On the other hand, SOC with the intermediate strength may still be strong enough to impose a competing, singlet ground state or an angular momentum J=0 state. Novel magnetic states may thus emerge when the singlet-triplet splitting (0.05-0.20 eV) becomes comparable to exchange interactions (0.05-0.10 eV) and/or non-cubic crystal fields[25-27]. This is evidenced in a recent study of materials containing $5d^4$ ions [28].

Up until now, no physical and structural properties of $Na_2RuO_3$ have been investigated but a few experimental and theoretical studies of polycrystalline $Li_2RuO_3$ have been reported in recent years [29-32]. In essence, polycrystalline $Li_2RuO_3$ undergoes a structural phase transition near $T_D$=540 K that features a change of space group from C2/m (No. 12) at high temperatures to $P2_1/m$ (No. 11) at low temperatures. The low-temperature phase adopts a strongly distorted honeycomb lattice, which prompts a simultaneous dimerization that results in a singlet ground state [29]. The observation of dimerized zigzag chains has recently stimulated more investigations of $Li_2RuO_3$[30-32], in which the dimerization is attributed to orbital ordering [29], creation of valence bond crystal [30] andJahn-Teller distortions [31], respectively. It is noted that all reported experimental results were culled from polycrystalline $Li_2RuO_3$[29, 31, 32].

Here we report structural, magnetic, and thermal properties of single-crystal $Li_2RuO_3$ and $Na_2RuO_3$.The newly found $Na_2RuO_3$with space group C2/mfeatures a nearly ideal honeycomb lattice and orders antiferromagnetically below 30 K. It may serve as a reference for almost perfect honeycomb symmetry.On the other hand, single-crystal $Li_2RuO_3$ adopts a less ideal honeycomb lattice with either C2/m or more distorted $P2_1/m$ below 300 K but both phases exhibit a well-defined, though different, magnetic state, which sharply contrasts with the singlet ground state due to dimerization observed in polycrystalline $Li_2RuO_3$[28]. This work producesa phase diagramthat uncovers a direct correlation between the ground state and basal-planedistortions or lattice-tuned magnetismin all honeycomb ruthenates studied.(Both $Li_2RuO_3$ and $Na_2RuO_3$ are highly insulating; their transport properties are not included in this paper.)

*Crystal Structures* Single-crystals of $Li_2RuO_3$ and $Na_2RuO_3$ were synthesized using the self-flux method, which is described elsewhere **[17]**. For synthesis of single-crystal $Li_2RuO_3$the mixed chemicals were first heated up to 1250 $^o$C and then cooled to 900 $^o$C at 2 $^o$C/hour and finally room temperature at 50 $^o$C/hour. In contrast, the polycrystalline $Li_2RuO_3$ was synthesized at much lower temperature of 950 $^o$C.Thedifferent synthesis conditionsmay have important implications for the ground state of $Li_2RuO_3$.For more experimental details,see Supplemental Material **[33]**. Crystal structures on which the ground state so sensitively hinges require a close examination.**Table 1**includesthe lattice parameters of single-crystals $Li_2RuO_3$and $Na_2RuO_3$ as well as those of polycrystalline $Li_2RuO_3$ and iridate counterparts for contrast and comparison. For the sake of discussion, single-crystal $Li_2RuO_3$ with C2/m and P2$_1$/m are labeled as $Li_2RuO_3$ (C)and $Li_2RuO_3$ (P), respectively. A major distinction between $Li_2RuO_3$ (C)and $Li_2RuO_3$ (P) is the number of unequal Ru-Ru bond distances, which measures distortions that in turn dictate theground state. $Li_2RuO_3$ (C)features two bond distances, or a long and short one, $L_l$ and $L_s$, respectively,whereas $Li_2RuO_3$ (P)has three bond distances, i.e., $L_l$, $L_s$,and a medium bond distance, $L_m$. The basal-plane distortion is characterized by the bond difference ratio defined as $(L_l-L_s)/L_s$, which is shown in **Table 1**,**Figs 1a** and **1b**. In general,honeycomb lattices with C2/mtend to have a larger$a$-axis lattice parameter and smaller ratio b/a($\sim\sqrt{3}$) than those with P2$_1$/m, thus less distorted.**Figs.1c** and **1d**demonstrate the lattice parameters of single-crystal and polycrystalline samples as a function of temperature. As seen, no structural transition is discerned in the single crystals studied for the temperature range measured. *In short, the structural differences between the polycrystalline Li$_2$RuO$_3$ and Li$_2$RuO$_3$ (C) or Li$_2$RuO$_3$ (P) are distinguished by the different space groups or by the difference in $(L_l-L_s)/L_s$. It is clear that Li$_2$RuO$_3$ (P) is more distorted than Li$_2$RuO$_3$ (C) but much less distorted than the polycrystalline sample despite the same space group shared by both* (**Table 1**).

**Table 1. Structural comparison between the honeycomb lattices at 100 K**

| Compound | Space Group | a (Å) | b (Å) | b/a | $(L_l-L_s)/L_s$ |
|---|---|---|---|---|---|
| Li$_2$RuO$_3$ (Powder)* | P2$_1$/m | 4.9210(2) | 8.7829(2) | 1.785 | 18.6% |
| Li$_2$RuO$_3$ (P) | P2$_1$/m | 4.963(3) | 8.766(6) | 1.766 | 10.1 % |
| Li$_2$RuO$_3$ (C) | C2/m | 5.021(4) | 8.755(6) | 1.744 | 2.1 % |
| Na$_2$RuO$_3$ | C2/m | 5.346(1) | 9.255(2) | 1.731 | 0.17 % |
| (Li$_{0.9}$Na$_{0.1}$)$_2$IrO$_3$ | C2/m | 5.186(1) | 8.964(2) | 1.728 | 0.6 % |
| Na$_2$IrO$_3$ | C2/m | 5.319(1) | 9.215(2) | 1.732 | 0.14 % |

*\* Taken at 300 K*

*Physical Properties* Na$_2$RuO$_3$ exhibits a sharp antiferromagnetic (AFM) transition at T$_N$=30 K, as shown in **Fig. 2a**. The magnetic anisotropy leads to a stronger out-of-plane magnetic susceptibility $\chi_\perp$ than in-plane magnetic susceptibility $\chi_\parallel$. The linearity illustrated in 1/$\Delta\chi_\parallel$ (right scale in **Fig. 2a**) indicates that the data fit well with the Curie-Weiss law for 100 < T < 350 K, and yield the Curie-Weiss temperature $\theta_{CW}$ = -137 K and effective moment $\mu_{eff}$ = 2.45 $\mu_B$/Ru (**Table 2**). The frustration parameter defined as FP = |$\theta_{CW}$|/T$_N$ is estimated to be 4.6. This value suggests a presence of modest frustration, comparable to that for its iridate counterpart.

The magnetic ordering is confirmed by the specific heat C(T) (**Fig. 2b**). However, an additional peak at T$_{N2}$ = 26 K that is absent in $\chi$(T) is also seen in C(T). This behavior, which is reproducible, is remarkably similar to that observed in Na$_2$IrO$_3$ where an additional, weaker anomaly in C(T) is discerned at T* = 21 K that is followed by the zigzag order at $T_N$ = 18 K **[15, 17]**. This two-step transition is discussed in context of the Kitaev-Heisenberg model on the hexagonal lattice **[34]**. A similar argument could be applied to $Na_2RuO_3$ although the origin of this magnetic behavior needs to be further investigated.The C(T) data also indicatethat the entropy removal due to the two-step magnetic transition is small, less than 10% of Rln3 expected for an S=1 magnet. This implies that the magnetic ordering may not be fully developedperhaps in part because of the tendency of SOC to impose a singlet state. Application of magnetic field up to 14 T causes no visible changes in both C(T,H) and $\chi$(T,H).

**Table 2. Physical parameters of the *single-crystal* honeycomb lattices**

| Compound | $T_N$ (K) | $\theta_{CW}$ (K) | FP | $\mu_{eff}$ ($\mu_B$/Ru or Ir) |
|---|---|---|---|---|
| $Li_2RuO_3$ (P) | ~ 5 | -58 | 11.6 | 1.46 |
| $Li_2RuO_3$ (C) | 9 | -112 | 12.4 | 2.77 |
| $Na_2RuO_3$ | 30 | -137 | 4.6 | 2.45 |
| $(Li_{0.9}Na_{0.1})_2IrO_3$ | 7 | -18 | 2.6 | 1.95 |
| $Na_2IrO_3$ | 18 | -119 | 6.6 | 1.76 |

*FP stands for frustration parameter*

The magnetic properties of both single-crystal $Li_2RuO_3$(C) and $Li_2RuO_3$(P) are examined for 1.7< T < 900 K. Neither shows the singlet ground state observed in the polycrystalline $Li_2RuO_3$. Instead, $Li_2RuO_3$(C) displays paramagnetic behavior at T > 20 K with the magnetic susceptibility $\chi$ following the Curie-Weiss law for 20 K < T ≤ 750 K (**Fig.3a**). Data fits to the Curie-Weiss law yield an effective moment $\mu_{eff}$ = 2.77 $\mu_B$/Ru, consistent with that expected for an S=1 system, and a Curie-Weiss temperature $\theta_{CW}$=-112 K. A signature for a long-range order near $T_N$=9K is evident in both $\chi$(T) and C(T)(**Fig. 3b**).A largefrustration parameter, FP = $|\theta_{CW}|/T_N$ = 12.4 suggests the presence of significant frustration (**Table 2**). Indeed, the two unequal Ru-Ru bondsmay favor a formation of zigzag chains along the a-axis (see schematic in the inset of **Fig. 4**) as the inter-chain interaction is weakdue to the long Ru-Ru bond $L_l$. Therefore, nomagnetic orderingoccursuntil below $T_N$=9 K when three-dimensional correlations are established.

For more distorted $Li_2RuO_3$ (P), a magnetically ordered state also takes placebut at a lower temperature, $T_N$=4 K (**Fig. 3c** and **3d**). Remarkably, the magnetic anisotropy is much stronger, and the magnitude of $\chi_\perp$ is significantly larger than that in $Li_2RuO_3$ (C), implying the importance of SOC. However, the temperature dependence of $\chi$ at high temperatures is much weaker than that for $Li_2RuO_3$ (C). The results suggest that $Li_2RuO_3$ (P) is "half-way" to dimerizationas the lattice is more similarto that of the polycrystalline sample; the magnetic state eventually prevails below $T_N$=4 K because$Li_2RuO_3$(P) is after all not as distorted as the polycrystalline$Li_2RuO_3$.

*Computational Results* OurLDA (local density approximation) calculations using the LMTO (linearized muffit-tin orbitals) method **[35]** andWannier function projection method **[36]** show that the crystal-field splitting in the Ru $t_{2g}$ shell does not exceed 70 meV, indicating that the comparable SOC may play a significant role. However, the off-diagonal matrix elements of the Hamiltonian, hopping parameters, are even larger, ~ 200 meV, which is strong enough to form the quasi-molecular orbitals (QMOs) similar to those in $Na_2IrO_3$ where QMOsinvolve six Iratoms arranged in a hexagon and each Ir atom belongs to three differentQMOs, which dominate the formation of electronic structure**[13]**. The results of the optimization of the crystal structure performed in the GGA (generalized gradient approximation) calculations using the pseudopotential method **[37]** indicate that the nearly ideal honeycomb $Na_2RuO_3$ indeed corresponds to a minimum of the total energy for an AFM state. In addition, our GGA+U calculations show that a relatively small on-site Coulomb repulsion U ~ 1.5 eV is sufficient to suppress the dimerization observed in polycrystalline $Li_2RuO_3$. The band structure of single-crystal $Li_2RuO_3$ strongly differs from that of both $Na_2RuO_3$ and $Na_2IrO_3$ on the LDA level (**see SM Fig. 2[33]**); and consequently, there is no sign of the QMOs. According to a recent study **[30],** when one of QMOs (of E2u symmetry) is half-filled, the corresponding instability may induce the Jahn-Teller distortions (JTDs) that in turn lead to the dimerization. In less distorted single-crystal $Li_2RuO_3$, no sign of the JTDs is seen since the formation of the zigzag chains effectively removes the orbital degeneracy or JTDs. Therefore the zigzag chains constitute an alternative state to the dimerization when the JTDs are absent. However, both the zigzag chains and dimerized lattice cost certain elastic energy that tends to stabilize uniform structure, and the prevailing state sensitively depends on details of the band structure and bulk modulus of the system (see **SM[33]** for details).

Indeed, all relevant energies vigorously compete and critically bias their mutual competition to stabilize ground states. This explains that there exist nearly degenerate states in these materials, and the prevailing ground state critically depends on details of the structure, as illustrated in **Fig. 4**. The magnetic ordering systematically decreases with increasing $(L_l-L_s)/L_s$ and eventually vanishes at a critical value where the dimerization emerges, leading to the singlet ground state observed in polycrystalline $Li_2RuO_3$. All results strongly indicate a direct correlation between the ground state and basal-plan distortions. The newly found $Na_2RuO_3$ provides a reference for almost perfect honeycomb symmetry.

The absence of the dimerization in single-crystal $Li_2RuO_3$ cannot be due to either impurity or quality of the single crystals. In fact, the singlet ground state is unusually resilient to heavy impurity doping and is even enhanced by 5% Na doping (**see Fig. 3 in SM [33]**) andsurvives up to 50% Ir substitution for Ru in the polycrystalline samples **[31].** It is likely that the difference between the two forms of $Li_2RuO_3$ arises from different synthesis conditions, as discussed above, which might cause different degrees of site-disorder in the honeycomb network due to the similar ionic radius of Li and Ru, and/or slightly different stoichiometry (e.g. oxygen content) (see **SM [33]**). Hence, this work does not rule out the possibility that single-crystal $Li_2RuO_3$ having the same structural distortions and singlet ground state as polycrystalline $Li_2RuO_3$ may eventually form under certain synthesis conditions.

Thework also offers the following general observations. Both $Li_2RuO_3$ and $Li_2IrO_3$ are more structurally distorted and behave with more complexities than their Na counterparts.SOC is expected to impose a J=0 state for $Ru^{4+}(4d^4)$ ions (anda $J_{eff}$=1/2 state for $Ir^{4+}(5d^5)$ ions) butthe observed magnetic states in the honeycomb ruthenates as in many other ruthenates **[25]** indicate that SOC is not sufficient to induce a J=0 state. It is intriguing thatall honeycomb ruthenates and iridatesmagnetically order in a similar temperature range (see **SM-Fig. 4 [33]**)despite the different role of SOC in them.

*Acknowledgements* GC is thankful to Drs. Natalie Perkins and Y. B. Kim for discussions. SS and DK are grateful to Drs. Igor Mazin, Harald Jeschke, Roser Valenti and Je-Geun Park. This work was supported by the National Science Foundation via Grants Nos. DMR-0856234, DMR-1265162, DMR-1056536 (RKK), Russian Science Foundation via RSCF 14-12-00306 (SVS), German project FOR 1346, Cologne University via German excellence initiative (DK), DOE BES Office of Scientific User Facilities (FY) andChina Scholarship Council (JCW).

*Corresponding author; cao@uky.edu

**Captions**

**Fig. 1.** Diffraction images in the *(h0l)* plane of the single crystal $Li_2RuO_3$ with space group **(a)** $P2_1/m$ and **(b)** $C2/m$. **Insets**: The corresponding honeycomb lattice and Ru-Ru bond distances. The temperature dependence of **(c)** the *a*-axis and **(d)** the ratio *b/a* from our single crystal $P2_1/m$ phase (blue), $C2/m$ phase (purple), powder samples (red star), and powder data from Ref.28 (black circles). Note that the sharp diffraction pattern clearly indicates the high quality of the single-crystal $Li_2RuO_3$.

**Fig. 2.** *Single-crystal $Na_2RuO_3$*: **(a)** The temperature dependence of the magnetic susceptibility for the basal plane $\chi_{\square\square}(T)$ and out-of-plane $\chi_{\perp}(T)$ for single-crystal $Na_2RuO_3$; Right scale: $1/\Delta\chi_{\square\square}$ where $\Delta\chi = \chi - \chi_\square$ and $\chi_\square$ is the temperature-independent contribution to $\chi$. **(b)** The temperature dependence of the specific heat $C(T)$ and $\chi_{\perp}(T)$ (right scale).

**Fig. 3.** *Single-crystal $Li_2RuO_3$ (C)*: The temperature dependence of **(a)** the magnetic susceptibility $\chi_{\square\square}(T)$ and $\chi_{\perp}(T)$ and $1/\Delta\chi_{\perp}$ (right scale) for $1.7 < T < 850$ K and **(b)** the specific heat $C(T)$ and $\chi_{|\square}(T)$ and $\chi_{\perp}(T)$ (right scale) at low T. *Single-crystal $Li_2RuO_3$ (P)*: The temperature dependence of **(c)** $\chi_{\square\square}(T)$ and $\chi_{\perp}(T)$ and $1/\Delta\chi_{\perp}$ (Inset) and **(d)** $\chi_{|\square}(T)$ and $\chi_{\perp}(T)$ and $d\chi_{\perp}/dT$ (right scale) at low T.

**Fig. 4.** The Neèl temperature $T_N$ as a function of the bond distance ratio $(L_l-L_s)/L_s$ for all honeycomb ruthenates. Inset: a schematic of the honeycomb lattice featuring $L_l$

and L_s.

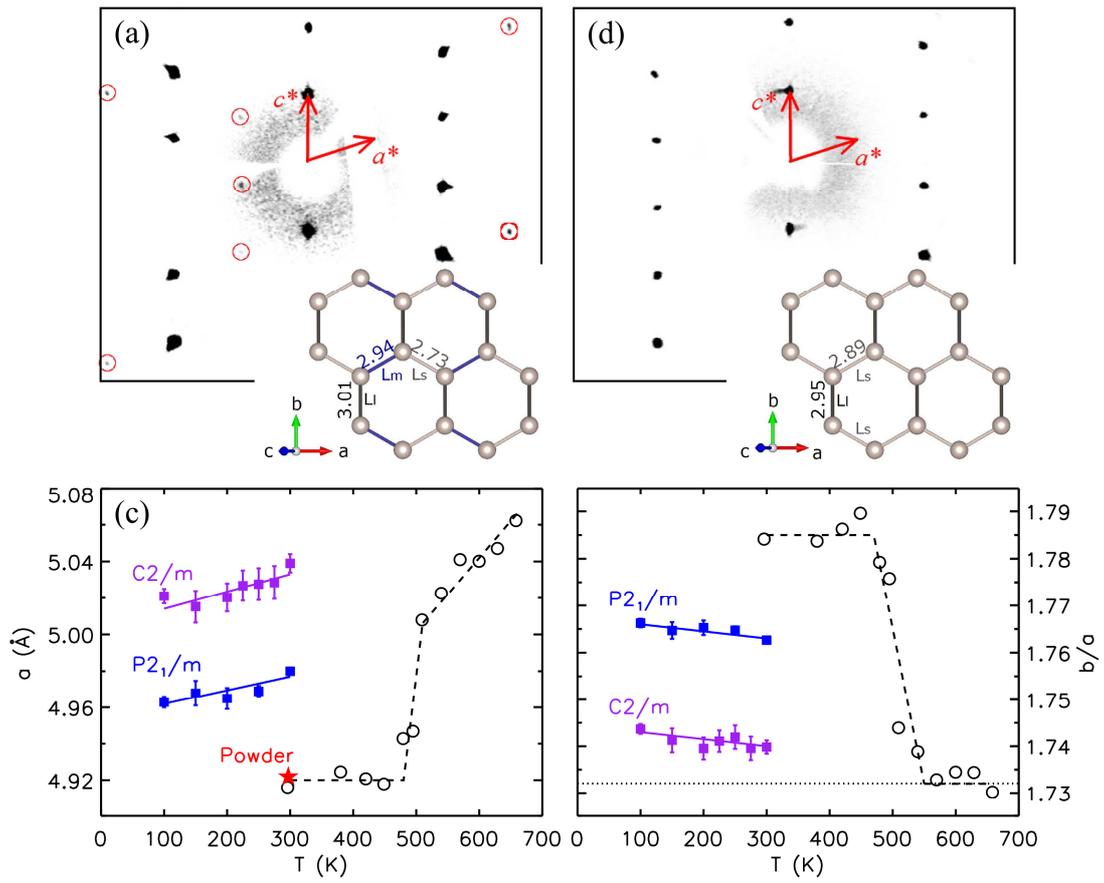

Fig. 1

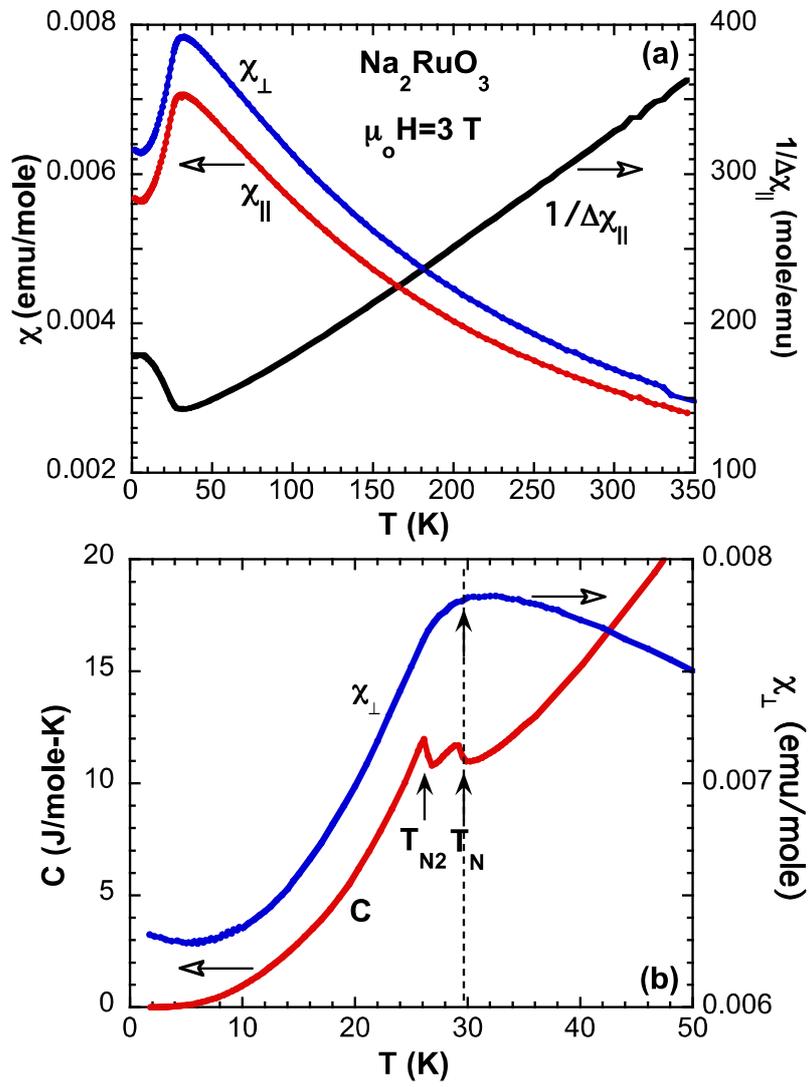

Fig. 2

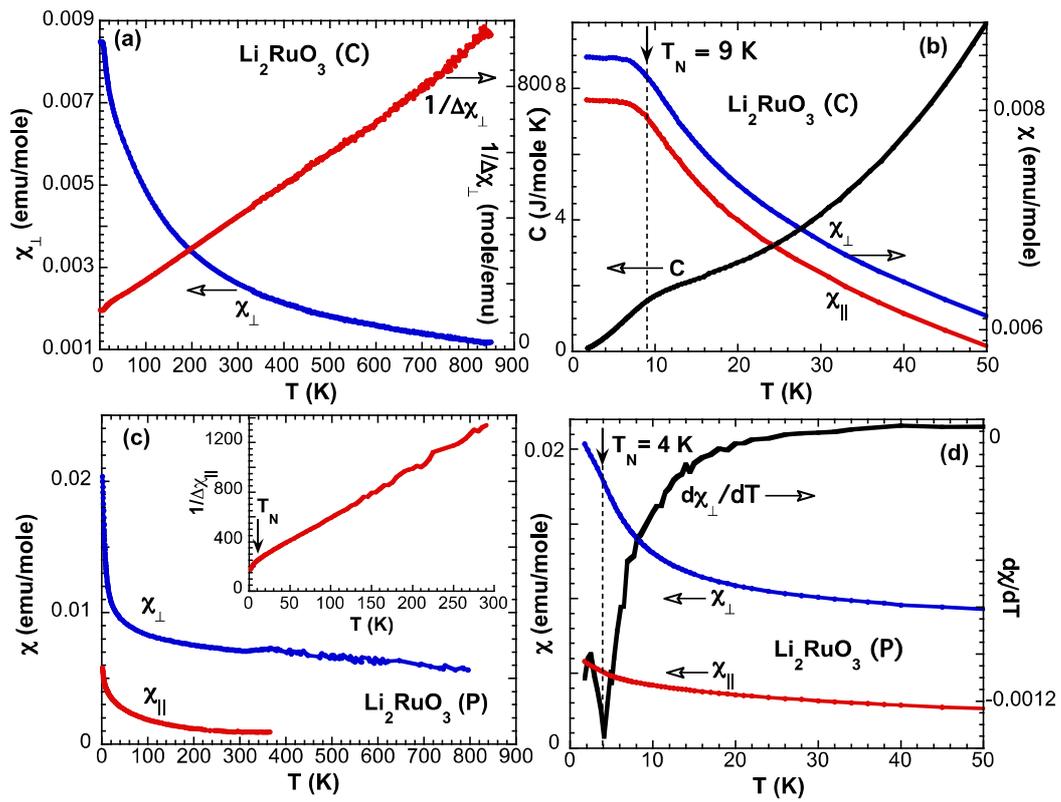

Fig.3

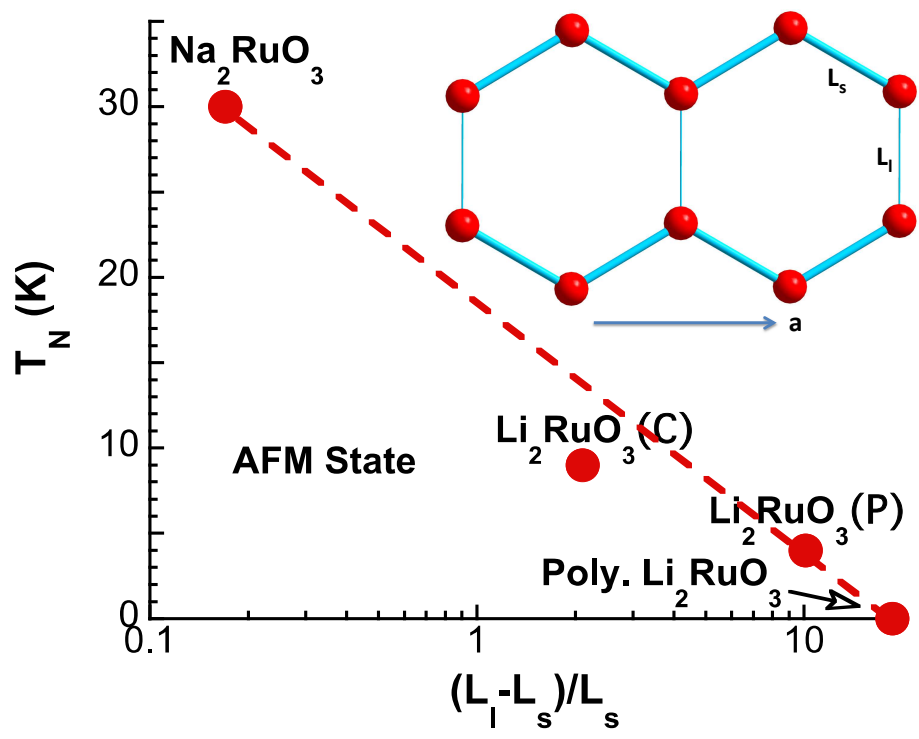

Fig. 4